\documentclass[12pt]{article}

\usepackage{soul}
\usepackage{newtxtext,newtxmath}
\usepackage{float}
\usepackage{graphicx}
\usepackage{epstopdf}

\usepackage[skip=10pt]{caption}
\usepackage[letterpaper,margin=0.8in]{geometry}

\renewenvironment{abstract}
	{\quotation}
	{\endquotation}

\date{}

\makeatletter
\renewcommand{\fnum@figure}{\textbf{Fig. \thefigure}}
\renewcommand{\fnum@table}{\textbf{Table \thetable}}
\makeatother

\usepackage{scicite}

\def\scititle{Pinned Ad-colloids Disfavors Nucleation in Colloidal Vapor Deposition}
\title{\scititle}

\author{
	Noman Hanif Barbhuiya$^{1}$,
	Pritam K. Mohanty$^{1}$,
	Saikat Mondal$^{2}$,
        Aminul Hussain$^{1}$, \and
        Adhip Agarwala$^{2}$,
        Chandan K. Mishra$^{1\ast}$\and
	\small$^{1}$Department of Physics, Indian Institute of Technology Gandhinagar, Palaj, Gandhinagar, 382055, Gujarat, India.\and
	\small$^{2}$Department of Physics, Indian Institute of Technology Kanpur, Kalyanpur, Kanpur, 208016, Uttar Pradesh, India.\and
        \small$^\ast$Corresponding author. Email: chandan.mishra@iitgn.ac.in
	}

\begin{document} 

\maketitle

\begin{abstract} \bfseries \boldmath

Crystallization through vapor deposition is ubiquitous, and is inevitably influenced by impurities, which often impact the local structure. Interestingly, the effect of immobilizing some of the depositing particles themselves, which would still preserve local structural symmetry, remains largely unexplored. Herein, we perform colloidal vapor deposition on a substrate with a few pinned ad-colloids, termed ``mobility impurities". Through thermodynamic and kinematic measurements, we demonstrate that these pinned ad-colloids, even though they share identical geometry and interaction with depositing particles, are disfavored as nucleation centers. We reveal that entropic contributions, rather than energetic ones, govern nucleation physics in the presence of mobility impurities. Moreover, tuning the mobility of colloids on the substrate adjusts the nucleation likelihood at pinned sites. In later stages of growth, pinning induces mode localization and alters the thin film's vibrational spectrum. Our work, thus, underscores the potential of strategically incorporating mobility impurities to engineer material properties.
\end{abstract}

\subsection*{INTRODUCTION}
The introduction of impurities, whether intentional or otherwise, during the initial stages of material fabrication holds considerable sway over a spectrum of material properties and their ensuing performance, such as structure \cite{granasy2003growth, musevic2006two}, mechanical strength \cite{zhong1993computer, kermode2013macroscopic} and transport properties \cite{delattre2009noisy, mocatta2011heavily}. These impurities, often extrinsic, differ from the primary constituents of the material in terms of one or more physical and chemical attributes and exert their influence by modifying the free energy landscape, thereby altering nucleation and growth mechanisms of the materials \cite{cacciuto2004onset, navrotsky2004energetic, elhadj2006role, de2005colloidal, allahyarov2015crystallization}. For example, in processes akin to vapor deposition, the presence of impurities can lead to diverse outcomes: they might lower the interfacial barriers to crystallization \cite{cacciuto2004onset}, and promote smoother film formation with reduced defect density \cite{ma2021immobilized, hong2020chemical}, regulate the rate of nucleation and the morphology of growing islands \cite{granasy2003growth}, or impede crystal growth altogether \cite{ma2020antagonistic, de2005colloidal}. However, a high density of immobile impurities on the substrate frustrates crystal nucleation \cite{katsuno2011effect} and would lead to a structurally disordered phase.

A thorough understanding of the precise influence of impurities requires disentangling multiple complex processes at play during these early stages of material fabrication. These include the shape and amplitude of the diffusion barrier of impurities on the substrate relative to the primary constituents \cite{kellogg1994direct, cacciuto2004onset, de2005colloidal, allahyarov2015crystallization}, as well as the nature of their interactions with the primary constituents compared to the interactions between the primary constituents themselves \cite{musevic2006two, liu1995submonolayer, cacciuto2004onset}. A clear microscopic understanding, therefore, requires a systematic and controlled isolation of these factors to elucidate their individual impacts. Interestingly, colloidal vapour deposition offers an excellent experimental platform to systematically isolate and investigate the role of each of these factors arising due to the presence of impurities on the physics of nucleation and growth of materials \cite{li2016assembly, manoharan2015colloidal}. 

Here, we uncover the role of contrasting diffusion barriers of impurities and the primary constituents (ad-colloids) on the substrate in colloidal vapor deposition while keeping all other attributes between them to be identical. To achieve this, we randomly immobilize a tiny fraction of depositing colloids themselves on the substrate prior to starting the deposition experiments, which we term as ``mobility impurities". These mobility impurities would be chemically and physically identical to the depositing particles, sharing the same interaction potential, but differ only in their lack of mobility \cite{van2008colloidal}. This sole kinetic distinction in the impurity from the depositing particles presents a unique opportunity to isolate the entropic contribution to the nucleation of precursor clusters, a factor often masked by energetic and/or geometric considerations in systems with conventional impurities.

We employ video microscopy to observe the spatiotemporal evolution of clusters, encompassing their formation, disintegration, and growth at a single-particle level, both with and without pinned particles. By developing novel analytical tools to understand the experimental data, we find distinct thermodynamic and kinematic mechanisms governing the colloid aggregation based on the presence or absence of pinned colloids in precursor clusters. Supported by minimal theoretical models and molecular dynamics simulations, our experiments reveal that pinning decreases the entropic contributions to the free energy, rendering pinned sites {\it disfavorable} for nucleation (see supplementary Movie S1). Surprisingly, this entropic penalty increases with cluster size despite the presence of only a single pinned particle in a cluster. Even though the number of free particles increases with cluster size, the effect of pinning is amplified, not diminished. Moreover, the entropic constraint grows as the mobility of the depositing colloids on the substrate decreases, providing a mechanism to control nucleation propensity at pinned sites. Thus, our work provides direct evidence that the physics of nucleation in the presence of these mobility impurities in colloidal vapor deposition are primarily entropy-driven rather than those by energetic considerations.

\subsection*{RESULTS}

\subsubsection*{Vapor deposition experiments}
\begin{figure}[hbt!]
\centering
\includegraphics[width=1.0\textwidth]{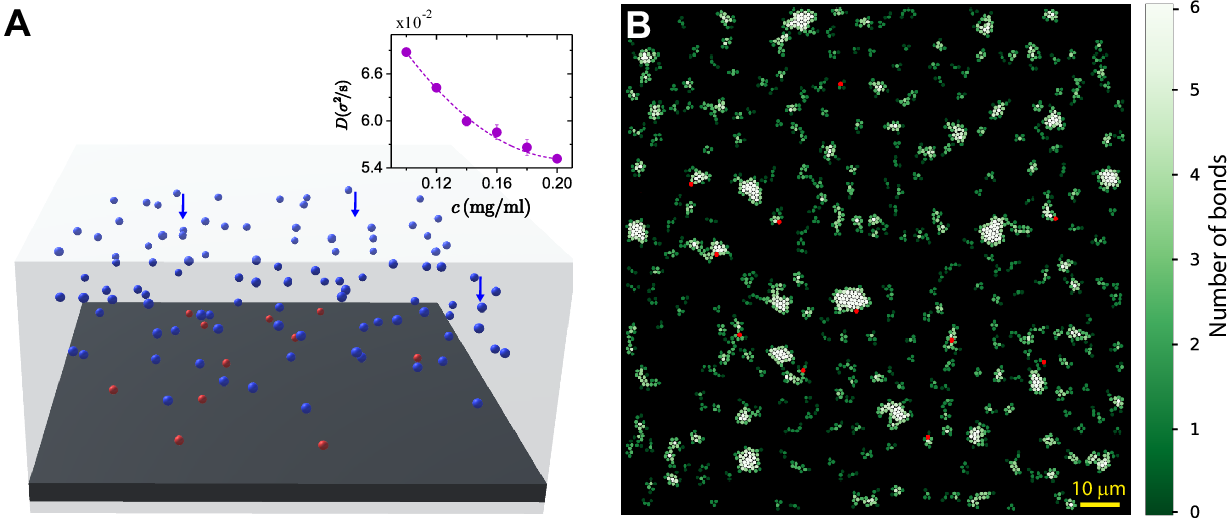}
\caption{\textbf{Experimental setup and characterization.} (\textbf{A}) Schematic of the experimental set-up showing pinned colloidal particles in red and the sedimenting ones in blue. Inset shows the surface diffusivity of single colloids, $D$ (in the units of $\sigma^2/$s), versus $c$. The dashed curve is a guide to the eye. Error bars for $0.12\leq c \leq 0.18$ mg/ml represent the standard error of the mean from two independent experiments. For $c = 0.20$ mg/ml, the error bars correspond to the fitting error. (\textbf{B}) Snapshot of the field-of-view at monolayer coverage, $\Theta \sim 0.17$, at $c = 0.12$ mg/ml. The particles have been color-coded based on the number of bonds they form in their nearest-neighbor shell. Pinned particles are shown in red.} 
\label{fig:Figure1}
\end{figure}

Figure \ref{fig:Figure1}A schematically illustrates our experimental setup for vapor deposition. Colloidal particles of diameter $\sigma \sim 1.0$ $\mu$m were allowed to sediment onto a glass substrate under gravity at a steady flux of $F = (5.1 \pm 0.1)\times 10^{-5}$ monolayers/s (see Materials and Methods). The sedimenting colloids, upon reaching the substrate, diffuse on the surface with diffusivity, $D \sim 6 \times 10^{-2} \sigma^2/s$ (inset to Fig. \ref{fig:Figure1}A \& see Supplementary Note S1). Thus, the sole control parameter in vapor deposition experiments, $D/F$, approaches the practically achievable upper limit in colloidal deposition experiments, which in our case turns out to be $D/F \sim 10^3$ \cite{ganapathy2010direct}. Prior to the start of the experiment, the glass substrate was featured with randomly positioned pinned particles that matched the sedimenting colloids in shape, size, and inter-particle interactions but differed only in mobility. In all our experiments, the average distance between any two closest pinned sites, $L_p$, was kept sufficiently large to ensure nucleation events at one pinned site are independent of others (see Materials and Methods). Moreover, since $L_p \sim 22\sigma$ was significantly larger than the mean diffusion length, $l \sim 6 \sigma$, of the depositing colloids on the substrate, nucleation events, whether involving pinned particles or not, were not only independent but occurred under identical experimental conditions. Of course, in the other extreme, where $l >> L_p$, mobility impurities would hinder crystal nucleation, resulting in disordered growth of the material \cite{katsuno2011effect}.

The depositing colloids, after sedimentation, eventually aggregate with neighboring colloids, either with or without a pinned particle. To minimize the desorption of sedimented colloids from the substrate, non-adsorbing depletants were added to the colloidal suspension, inducing short-range attractive forces between colloids and the substrate as well as between colloids themselves. The introduction of depletants serves another purpose in our experiments: increasing the concentration of depletant particles, $c$, decreases diffusivity of colloids on the substrate (inset to Fig. \ref{fig:Figure1}A, see Materials and Methods). A reduction in $D$ decreases mobility contrast between the pinned and free colloids, and thus, $c$ could be a potential control parameter for modulating the probability of nucleation at pinned sites.

Strikingly, regardless of the value of $c$, pinned sites were seldom observed to be at the center of the growing precursor clusters (Fig. \ref{fig:Figure1}B and Supplementary Fig. S1), indicating thermodynamics and kinematics of the nucleation mechanism may be altered in the presence of mobility impurities. To this end, we have developed novel analytic tools to directly measure these relevant parameters utilizing particle trajectories extracted from our video microscopy experiments. These tools enable us to uncover and contrast the underlying microscopic mechanisms for nucleation in the presence of mobility impurities.

\subsubsection*{Thermodynamics of precursor clusters}
For cluster nucleation to be favored at the pinned sites, the free energy difference between precursor clusters of a specific size $n$, comprising a pinned particle and those composed of only free particles must be negative, \textit{i.e.}, $\Delta F_{p-f}(n) = [\Delta U_{p-f} (n) - T\Delta S_{p-f}(n)] <  0$. Here, $\Delta U_{p-f} (n)$ and $\Delta S_{p-f}(n)$ are the differences in the internal energy and entropy between clusters of the same size $n$ formed with a pinned particle and without it, and $T$ is temperature. In our calculations utilizing experimental data, $\Delta U_{p-f} (n)$ and $\Delta S_{p-f}(n)$ are ensemble averaged over all the clusters of size $n$ appearing over the total experimental duration, $t_{tot} \sim 10^4$ s. However, note that primary contributions to the statistics arise predominantly from the early stages of film growth (Supplementary Fig. S2).

Given the steady low flux rate ($\sim 10^{-5}$ monolayer/s) and typical diffusivity of about $6 \times 10^{-2} \sigma^2/$s (inset to Fig. \ref{fig:Figure1}A), the nucleation processes at individual precursor clusters can be considered to be in a quasi-static equilibrium with the local environment. While nucleation is inherently a non-equilibrium dynamical process, concepts from equilibrium statistical mechanics are often crucial to building their understanding. Hence, we employed conventional thermodynamic variables such as $\Delta U_{p-f} (n)$ and $\Delta S_{p-f}(n)$ from the observations of the colloid dynamics to investigate any potential preferential nucleation at pinned sites.

\begin{figure}[h!tbp]
\centering
\includegraphics[width = 1.0\textwidth]{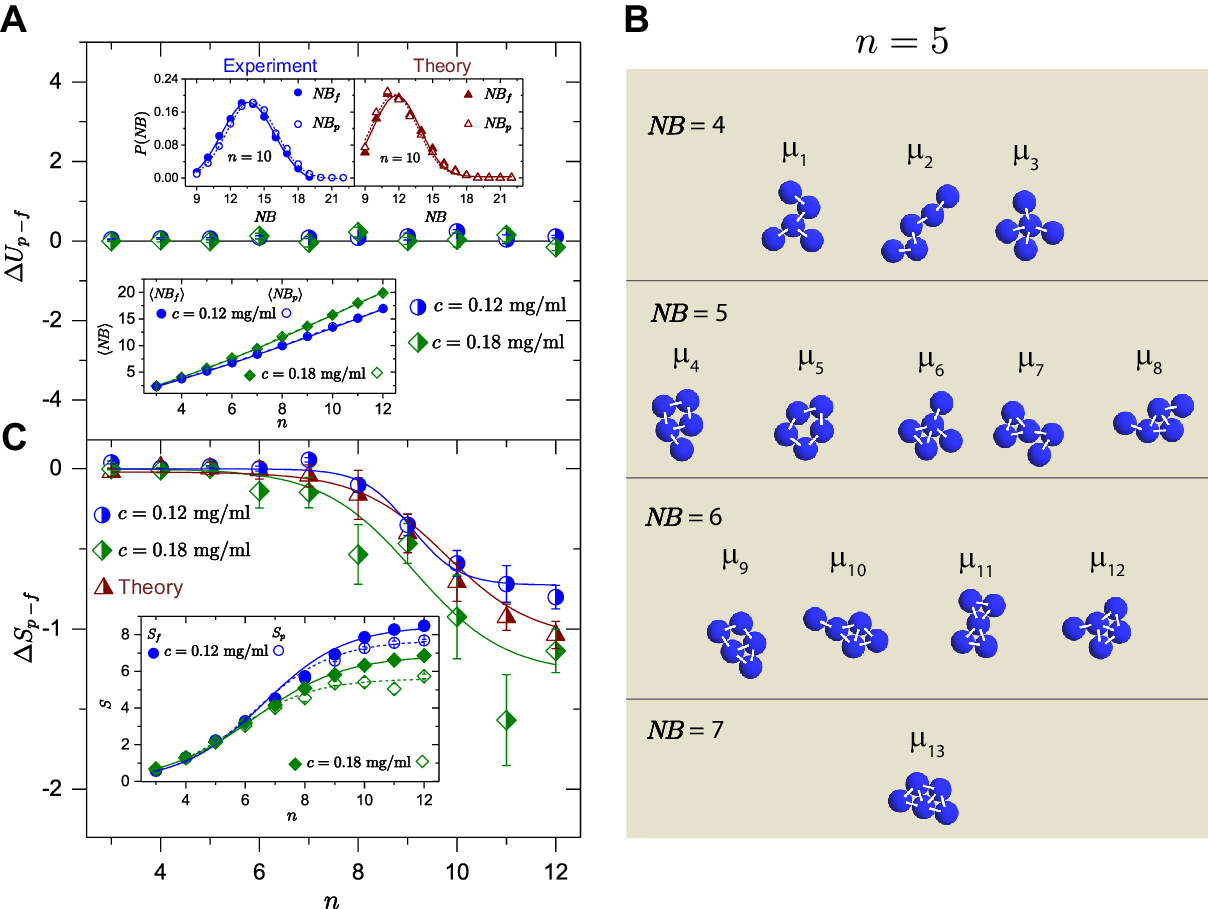}
\caption{\textbf{Internal energy and entropy of precursor clusters.} (\textbf{A}) $\Delta U_{p-f}(n)$ in the units of scaled $k_B T$ versus $n$ at different $c$. The top panel inset shows $P(N\!B)$ for $n =10-$particle clusters from experiment at $c = 0.12$ mg/ml (left panel) and theory (right panel). The bottom inset shows $\langle N\!B (n) \rangle$ versus $n$. Open and closed symbols represent data for clusters with and without a pinned particle, respectively. (\textbf{B}) Enumeration of experimentally observed topologically distinct coarse-grained microstates, $\{\mu_i\}$, for $n = 5$, with unique arrangements of bond connectivity for different total number of bonds, $N\!B$. (\textbf{C}) $\Delta S_{p-f}(n)$ in the units of $k_B$ versus $n$ for the same $c$ as in (\textbf{A}) and theoretical model (brown). The inset shows the entropy of clusters as a function of $n$. Open and closed symbols represent data for clusters with and without a pinned particle, respectively. For theoretical curves in (\textbf{A}) and (\textbf{C}), cluster distributions are obtained from steady-state configurations of a Markov matrix governing hard spheres hopping on a triangular lattice of 256 sites. About 10$^5$ configurations are taken for each of the particle densities between $0.1$ and $0.6$ in intervals of $0.01$, and for the pinning case, $1\%$ of sites are kept immobilized. Solid and dashed curves in (\textbf{A}) and (\textbf{C}) are guides to the eye. Error bars in data points represent the standard error of the mean from two different experiments at the same $c$, consisting of $\sim 10^4-10^5$ data points.}
\label{fig:Figure2}
\end{figure}

Due to the short-range nature of depletion-induced attraction, the internal energy of a cluster with a fixed size $n$ is proportional to the number of bonds, $N\!B (n)$, in the cluster  \cite{meng2010free}. Analyzing all the cluster configurations appearing over $t_{tot}$, we find $ P(N\!B)$ to be unimodal for all $n$, irrespective of whether a cluster contains a pinned particle or not (top left inset to Fig. \ref{fig:Figure2}A). 

Next, using $P(N\!B)$, we find the average number of bonds, $\langle N\!B(n) \rangle$, for each $n$ (bottom inset to Fig. \ref{fig:Figure2}A), and subsequently estimate $\Delta U_{p-f} (n)$ (Fig. \ref{fig:Figure2}A). As expected, for a fixed $c$, $\langle N\!B(n) \rangle$ increases with $n$. For a given $n$, $\langle N\!B(n) \rangle$ increases with $c$ due to increased inter-particle attraction favoring compact bond configurations. Interestingly, $\langle N\!B(n) \rangle$ exhibits similar trends for clusters with and without a pinned particle, yielding $\Delta U_{p-f} (n) \sim 0$ (Fig. \ref{fig:Figure2}A). Thus, from the perspective of internal energy, the inclusion of a pinned particle in precursor clusters may have no impact on their thermodynamic stability. Having discussed the role of internal energy, we now shift our focus to entropic contributions to free energy. 

When considering a cluster of size $n$, multiple unique bond connectivities are possible for a fixed $N\!B$, which we define as topologically distinct coarse-grained microstates, $\{\mu_i\}$ (Fig. \ref{fig:Figure2}B) \cite{perry2015two}. Thus, we can determine the entropic contributions by the identification and enumeration of these microstates. To catalog the unique microstates, we map all bond configurations for each $n$ onto their respective adjacency matrices, ${A_{n\times n}}$. Each element $a_{(l, m)}$ of this matrix indicates the presence (1) or absence (0) of a bond between particles $l$ and $m$. Subsequently, we identify the unique isomorphic bond configurations (microstates) \cite{juttner2018vf2++, cordella2001improved, hagberg2008exploring}, and obtain the probabilities ${p(\mu_i)}$ for each $\mu_i$ corresponding to a particular $n$ (see Supplementary Note S2 and Supplementary Fig. S3). The entropy of a cluster of size $n$ can then be defined as $S(n) = - k_B \sum_{i} {[p (\mu_i) \log(p(\mu_i))]}_n$. Here, $p(\mu_i)$ represents the probability of the $i^{\text{th}}$ unique microstate and $k_B$ is the Boltzmann constant. Unlike internal energy, the entropy for $n \geq 7$ exhibits a distinct profile between clusters featuring a pinned particle and those comprising only free particles (inset to Fig. \ref{fig:Figure2}C). For a given $c$, $\Delta S_{p-f} < 0$ (Fig. \ref{fig:Figure2}C), leading to $\Delta F_{p-f} > 0$, suggesting that pinned particles are disfavorable for nucleation. 

For a cluster of size $n$ containing a pinned particle, the number of free particles is $n-1$. As $n$ increases, both the number of free particles in the cluster and the total degrees of freedom increase, while the restricted degrees of freedom due to pinning remain constant. Therefore, one would expect the impact of mobility impurities to diminish as $n$ grows. However, our experiments suggest otherwise, indicating that pinning \textit{even a single particle} in precursor clusters significantly impacts its nucleation mechanism and the free energy landscape. Moreover, as evidenced by the variation of $\Delta S_{p-f} $ with $c$, the entropic constraints on clusters with a pinned particle increase with decreasing $D$ (Fig. \ref{fig:Figure2}C). Thus, $c$ can be adjusted to tune nucleation likelihood at pinned sites.

Remarkably, while our experiments, for the first time, uncover the role of mobility impurity and configurational entropy in colloidal vapor deposition with short-ranged depletion-induced attractions with a specific functional form \cite{asakura1958interaction}, the findings are expected to have broad applicability. The observed free energy penalty due to pinning arises solely from the reduced configurational space of the precursors containing it, without affecting their internal energies. To corroborate our arguments, we performed molecular dynamics simulations of vapor deposition using the LAMMPS package with two different interaction potentials: Lennard-Jones and Morse potentials \cite{LAMMPS}, commonly used to study diverse systems over varying lengthscales. The inferences drawn from these simulations mirror our experimental findings, demonstrating the universality of our study (see Supplementary Note S4 and Supplementary Fig. S4).

Guided by experimental phenomenology, we construct a minimal theoretical model. Modeling colloids as hard spheres hopping on a triangular lattice, the Markov matrix can be mapped to a ferromagnetic spin${-1/2}$ Heisenberg model \cite{feller68_prob, livi_Politi_2017}. The ground state of such a Hamiltonian, corresponding to the steady state of the Markov matrix, represents an equal probability distribution of all possible states of a fixed number of particles on the lattice, satisfying the hard sphere constraint. Pinning a particle thus corresponds to fixing a spin to $+1/2$ state on the particular site in the Heisenberg lattice Hamiltonian (see Materials and Methods).

By sampling typical configurations in the steady state ($\sim 10^5$), we isolate the cluster distributions for a fixed cluster size $n$, both with and without a pinned particle (see Supplementary Note S5). Consistent with the experimental findings, we observe no significant change in $P(N\!B)$ due to pinning of a particle (top right inset to Fig. \ref{fig:Figure2}A, and Supplementary Figs. S5 \& S6). This indicates that the bond probability distribution is primarily governed by lattice animal configurations rather than energetic considerations \cite{essam_1980, stauffer2018introduction, sykesI_1976}.  Furthermore, the configurational entropy of the clusters reveals that pinning a particle indeed reduces the entropy of the cluster (Fig. \ref{fig:Figure2}C and Supplementary Fig. S7), making them less favorable for nucleation. This also makes it plausible why pinned sites either remain as single entities or are predominantly located at the edge of the crystallite in our experiments (Fig. \ref{fig:Figure1}B, and Supplementary Fig. S1).

Interestingly, since we consistently observe that $\Delta F_{p-f} \sim 0$ for $n<7$ in our experiments, theoretical model and molecular dynamic simulations, it is crucial to determine whether the impact of mobility impurities in nucleation mechanisms diminishes with increasing $c$. At higher $c$, nucleation is expected to initiate at smaller cluster sizes compared to those at smaller $c$. Hence, we first attempt to identify the critical nuclei size, $n_c$, at all $c$ studied in this work, and for both the scenarios, with and without pinned particles. $n_c$ marks the threshold where cluster growth becomes favorable (Supplementary Fig. S8). To determine $n_c$, we track all clusters of size $n$, starting with $n=2$, with time and their subsequent most probable cluster size. For a given $c$, $n_c$ is defined as the smallest $n$ for which the subsequent most probable cluster size exceeds $n$. In addition, for all $n \geq n_c$, the next most probable cluster size must be greater than $n$.

We find that for clusters formed out of all free particles, $n_c$ can be determined for all $c$ and are within the range $4 \leq n_c \leq 13$ (Supplementary Fig. S9). However, surprisingly, with a pinned particle, the analysis for $n_c$ displays no discernible trend at any $c$ (Supplementary Fig. S8), indicating a significant disruption to the nucleation process of precursor clusters. The disruption in the nucleation process for clusters containing a pinned particle can be understood using thermodynamic arguments for clusters of size $n \geq 7$ (Fig. \ref{fig:Figure2}), which is the case for $c \leq 0.14$ mg/ml where $n_c > 7$ (Supplementary Fig. S9).

However, as $c$ increases, $n_c$ decreases, leading to a regime where $n_c < 7$ and the free energy difference is negligible ($\Delta F_{p-f} \sim 0$) between clusters with and without a pinned particle. In other words, the thermodynamic analyses suggest that clusters of size $n<7$, regardless of whether they contain a pinned particle or not, can access similar energy states with similar probabilities. However, the persistent difficulty in determining $n_c$ for clusters with a pinned particle for $c \geq 0.16$ mg/ml, implies that the equilibrium thermodynamic approach is insufficient to explain the observed nucleation mechanism. It is plausible that the kinetic nature of the impurity of the pinned particle introduces ``kinematic bottlenecks" in the configurational landscape of clusters containing a pinned particle. Consequently, the kinematic bottlenecks in clusters containing a pinned particle may impede their ability to efficiently navigate among various structural configurations, even when the thermodynamic driving force among them is comparable to that of clusters with all free particles. Essentially, the kinetic pathways connecting these states could differ, and that the dynamics of microstate transitions, and not just their equilibrium distribution, can play a crucial role \cite{doi:10.1126/science.1128649}. To investigate this, we now turn our attention to the dynamics of microstate transitions within the clusters of the same size.

\subsubsection*{Kinematics of precursor clusters}
During the initial stages of vapor deposition, colloids from the bulk join the precursor clusters on the substrate. The rate at which these clusters navigate their free energy landscape dictates the fate of precursor clusters and sets the stage for the stability and growth of larger clusters (and crystallites). For instance, in vapor-deposited molecular glasses, faster surface diffusion of molecules enhances their stability \cite{swallen2007organic, zhu2011surface}. Similarly here, identically sized clusters having equivalent thermodynamic stability, may still differ kinematically, influencing their suitability for ensuing growth. The rearrangement of particles within each cluster leads to exploration of various microstates. Consequently, the rate of exploration dictates the pace of navigating the free energy landscape, while the number of unique particles involved ensures thoroughness in this exploration.

\begin{figure}[h!tbp]
\centering
\includegraphics[width = 1.0\textwidth]{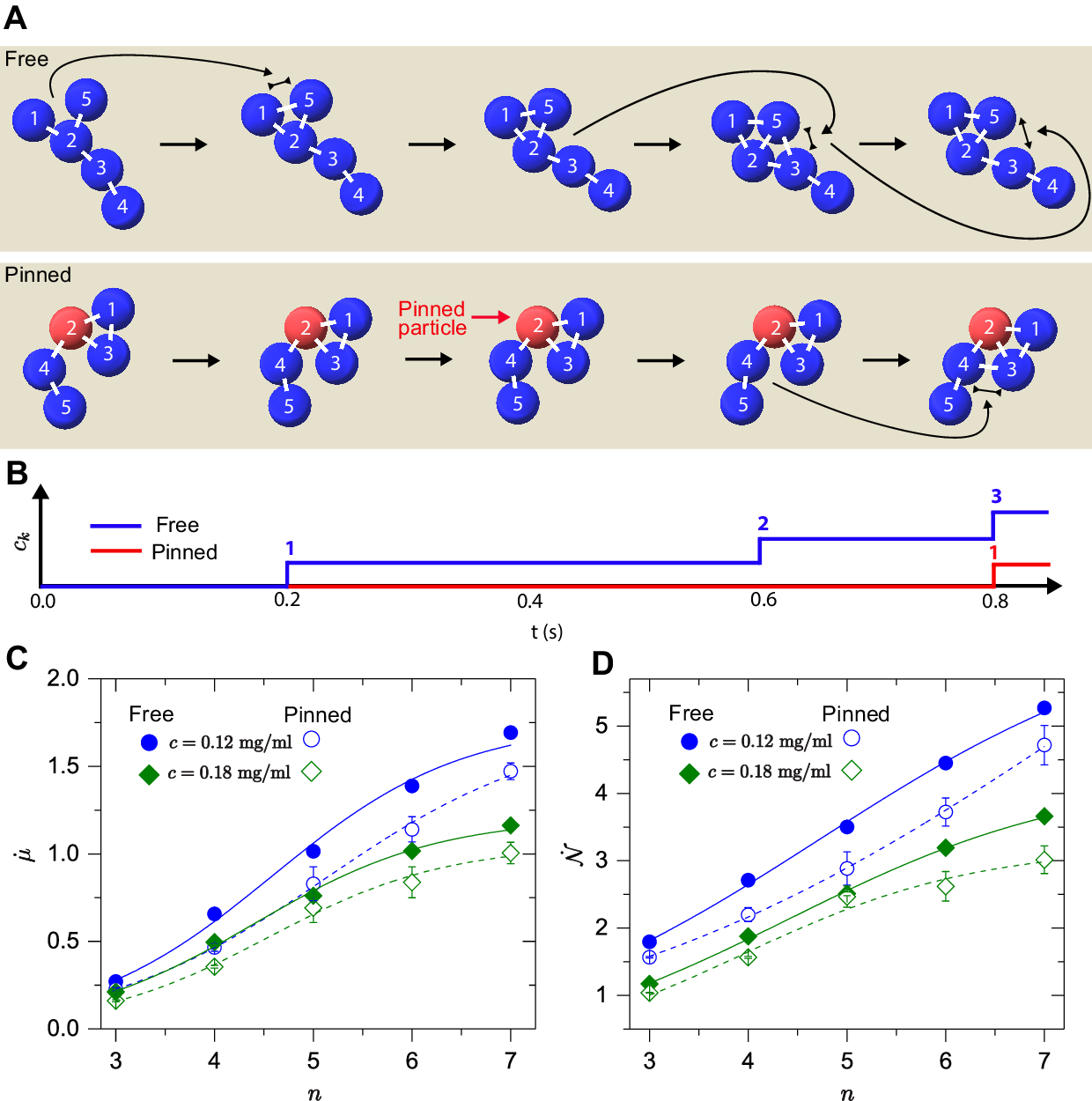}
\caption{\textbf{Kinematics of precursor clusters.} (\textbf{A}) Rendering from the experiment at $c = 0.12$ mg/ml to illustrate the change in microstate for a given $5-$particle cluster comprising of all free particles (top panel) and with one pinned particle (bottom panel) at $\Delta t = 0.2$ s. (\textbf{B}) Counter, $c_k$, tracking microstate transitions in a cluster with all free particles (blue line) and with one pinned particle (red line) for the temporal evolution of clusters shown in (\textbf{A}). (\textbf{C}) $\dot{\mu}(n)$ and (\textbf{D}) $\dot{\mathcal{N}}(n)$ versus $n$ for clusters (persisting for at least 0.5 s) with one pinned particle (open symbols) and with all free particles (solid symbols) at different $c$. Solid and dashed curves in (\textbf{C}) and (\textbf{D}) are guides to the eye. Error bars represent the standard error of the mean from two different experiments at the same $c$, consisting of $\sim 10^4 - 10^5$ data points.} 
\label{fig:Figure3}
\end{figure}

To quantify the rate, we track the temporal evolution of all clusters of size $n$ as they transition between different microstates, $\mu_k$ at a fixed time interval (Fig. \ref{fig:Figure3}A). For a cluster surviving for a time $t_k$ with cumulative instances of transitions $C_k$, we quantify the rate of accessing microstates $\dot{\mu}(n)= { \big [{\langle \frac{C_k}{t_k} \rangle}_{k} \big ]}_n$where ${\langle \cdot \rangle}_k$ represents ensemble average over all the clusters, $k$, of size $n$. Similarly, rate associated with total number of unique particles, $\mathcal{N}_k$, that participate in $\mu_k$ over $t_k$ is $\dot{\mathcal{N}}(n) = { \big [{\langle \frac{\mathcal{N}_k}{t_k} \rangle}_{k} \big ]}_n$. Note that the necessity for clusters to persist sufficiently longer in time precludes any such analysis for $n > 7$, the range already explored through thermodynamic arguments.

As anticipated, for a fixed $c$, both $\dot{\mu}(n)$ and $\dot{\mathcal{N}}(n)$ increases with $n$ (Figs. \ref{fig:Figure3}C and D, and Supplementary Fig. S10). Remarkably, while $\Delta F_{p-f} \sim 0$ for $n < 7$, $\dot{\mu}(n)$ and $\dot{\mathcal{N}}(n)$, are consistently lower for clusters containing a pinned particle (Fig. \ref{fig:Figure3}C and D, and  Supplementary Fig. S10). Thus, pinning a particle in clusters not only slows its traversal through the free energy landscape but also engages fewer unique particles in their exploration. This hampers the ability of clusters with a pinned particle to promptly attain a stable configuration, as evidenced by the difficulty in determining $n_c$ (Supplementary Fig. S8). Taken together, pinned particles fail to serve as preferred nucleation sites for the precursor clusters and the ensuing crystal growth (see supplementary Movie S1).

\subsubsection*{Spatial extent of vibrational modes in crystallites}
The formation of critical nuclei, albeit without a preference to nucleate at the pinned colloids, and the increase in the number density of the island, measured from the number of all clusters with size $n \geq 2$, signals the onset of the crystals growth regime (Supplementary Fig. S2). The island density continues to rise until (predominantly) newly sedimented colloids encounter a nucleating island within their mean diffusion length, $l$. Beyond this point, the clusters grow and then begin to coalesce, culminating in the formation of a thin film. While each of these processes is essential for a comprehensive understanding of the physics of nucleation and growth, our focus now shifts to the final stage of film growth, where sedimentation is nearly complete, and the islands have coalesced. This would allow us to investigate the impact of mobility impurities on the dynamical properties of the larger crystallites at large monolayer coverage, $\Theta$. 

In this regime, while individual particles on the crystallite periphery may still rearrange, the system has effectively reached equilibrium concerning the island-scale structures (Supplementary Fig. S11). Not surprisingly, the vibrational amplitude of colloids noticeably decreases in proximity to a pinned particle (Supplementary Fig. S11). This decrease suggests the localization of vibrational modes within the film, which we have quantified. However, given the challenges of locating crystallites with a pinned particle at their center and the presence of diffusing colloids in the second layer of the growing thin film, potentially disrupting analysis of vibrational modes, we adapted to a new experimental design (see Materials and Methods). Remarkably, within an experimental area of about $ 10^5 \sigma^2$, we only encountered a tiny fraction ($\sim 1\%$) of instances wherein crystallites contained a pinned particle away from their edges; the remainder were located at the periphery of the crystallite. This concurs with our earlier conclusion that pinned particles do not facilitate nucleation.

\begin{figure}[h!tbp]
\centering
\includegraphics[width = 1.0\textwidth]{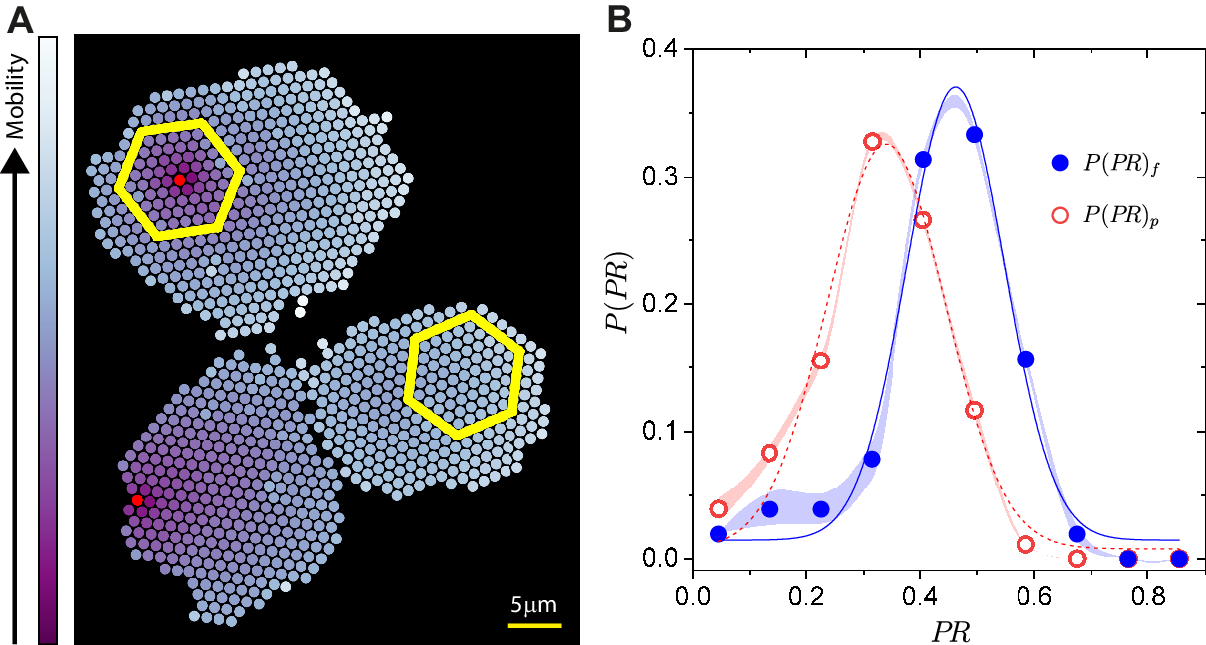}
\caption{\textbf{Localization of normal modes of vibrations.} (\textbf{A}) Snapshot of crystallites at $c = 0.12$ mg/ml showing sub-regions (yellow hexagon) utilized for analysis of localization of normal modes of vibrations. Free particles are color-coded based on their mobility, and pinned particles are rendered in red. The particle's mobility is calculated as the square root of the standard deviation of the particle's position, determined for and consequently averaged over all non-overlapping one-second slices of data. (\textbf{B}) Probability distribution of the participation ratio $P\!R$ for the sub-regions with a pinned particle (open symbols) and with all the free particles (solid symbols). The respective standard error of the means obtained from two different experiments with similar sub-regions, as shown in (\textbf{A}), are shown using ribbons. The dashed (red) and solid (blue) curves are Gaussian fits to the data.}
\label{fig:Figure4}
\end{figure}

To analyze the localization of the vibrational modes, we first calculate the normal modes of vibrations for two sub-regions: one with the pinned particle at the center and the other without it (Fig. \ref{fig:Figure4}A). This calculation, performed on the displacement covariance matrix, provides eigenvectors, $\textbf{e}_m$, and eigenfrequencies, $\omega_m$ (see Supplementary Note S6) \cite{chen2010low, melio2024soft}. Here, $m$ denotes the mode index. Next, we measure the participation ratio of the particles in the sub-region to each of the modes, ${[P\!R(\omega)]}_m$, that satisfies Debye scaling (Supplementary Fig. S12). ${[P\!R(\omega)]}_m = {\big (\sum_i^{N} (\textbf{e}_{m,i})^2 \big)}^2 / {\big (2N \sum_i^N (\textbf{e}_{m,i})^4 \big)}$, where $\textbf{e}_{m,i}$ is polarization vector of $i^{th}-$particle in mode $m$ and $N$ corresponds to the total number of particles in the sub-region \cite{chen2010low}. The probability distribution of the participation ratio, $P(P\!R)$, for the sub-region with a pinned particle peaks at a smaller value of $P\!R$ than those without it (Fig. \ref{fig:Figure4}B). Moreover, from the mean of the participation ratios, ${[\langle P\!R \rangle]}_p  < {[\langle P\!R \rangle]}_f$, which clearly indicates that the modes in crystallites with a pinned particle are more localized compared to crystallites with all free particles. Note, however, that the mobility impurities do not disturb the local structural symmetry of the crystallite and are otherwise statically indistinguishable from depositing particles (Supplementary Fig. S13). Thus, our work shows that strategic incorporation of such impurities can significantly impact and be used to engineer the mechanical properties of thin films and realize novel materials \cite{arsenault2007photonic, courty2005vibrational}.

\subsection*{DISCUSSION}
Our study unveils the critical influence of mobility impurities with unchanged inter-particle interaction potential on nucleation and growth of precursor clusters in colloidal vapor deposition. Through a combination of thermodynamic and kinematic measurements, using both experiments and associated theoretical insights, we demonstrate that pinned colloids fail to serve as preferred nucleation sites. Importantly, the nucleation probability at pinned sites can be further modulated by manipulating the mobility of free colloids on the substrate. Notably, the findings from our study on nucleation mechanisms in the presence of mobility impurities, in principle, are expected to extend over to other $D/F$ values as well, provided $l < L_p$. However, the disparity in nucleation behavior at a pinned site and without it is expected to be amplified at low $D/F$ values, where diffusive exploration is more restricted relative to the deposition rate. 

While our experiments reveal the crucial role of configurational entropy in governing the nucleation mechanism in colloidal vapor deposition in the presence of mobility impurities, molecular dynamics simulations under experimental-like conditions, using Lennard-Jones and Morse potentials, unambiguously establish the broader applicability of our findings. Moreover, in later stages of growth, even a single pinned site induces localization of vibrational modes in the thin film (large crystallite), potentially altering its mechanical stability. 

These findings collectively highlight the role of mobility impurities as a novel platform for influencing nucleation, growth, and material properties during the fabrication of colloidal superstructures, contributing to a broader understanding of mesoscale design principles \cite{arsenault2007photonic, courty2005vibrational}. Having established the impact of mobility impurities on the early nucleation stage, we recognize that the implications likely extend beyond the nucleation stage. Specifically, the influence of such impurities on intermediate-stage growth processes, such as cluster coalescence and Ostwald ripening, remains an open question and a promising avenue for future investigation. Moreover, while our investigation focused on short-range attractions relevant for clustering of colloids, the insights gained may offer valuable parallels to atomic and molecular systems, given the analogous epitaxial growth laws, albeit the underlying physics differ \cite{ganapathy2010direct}. Future studies could examine the influence of mobility impurities on the second and subsequent layers of crystal growth, as well as explore systems with long-range interactions, directional bonding, and the effects of pinning larger crystallites, which may be relevant to self-assembly in molecular systems

\subsection*{MATERIALS AND METHODS}
\label{sec:materials_and_methods}

\subsubsection*{Experiments to study nucleation and growth}
Colloids used in the study of nucleation and growth kinematics, reported in Figs. \ref{fig:Figure1}$-$\ref{fig:Figure3}, were synthesized using the modified Stober process  \cite{stober1968controlled, zhang2009hollow}. Post-synthesis, the particles were cleaned by repeated centrifugation and re-dispersion in water. The size of these particles was $\sigma \sim 1.0$ $\mu$m with size polydispersity $\sim 6 \%$. The experiments were conducted in glass flow cells (dimensions [length $\times$ width $\times$ height]: $\sim 22$ mm $\times$ $\sim 15$ mm $\times$ $\sim 1$ mm). A dilute suspension of spherical silica colloids, identical to the one used in the sedimentation experiment, was spin-coated onto a glass coverslip. This process resulted in featuring the glass substrate with randomly positioned pinned particles, which served as the bottom substrate of the flow cell. The spin-coating speed and the concentration of the sample used for spin-coating controlled the mean separation, $L_p$, between the pinned particles. In all our experiments, $L_p = (22.5 \pm 1.3) \sigma$ (mean $\pm$ standard deviation) measured over a field of view of size $\sim 114 \times 114 \mu\text{m}^2$ and across at least four such adjoining fields-of-view in all four directions. The pinned colloids were later identified from microscopy videos by analyzing the mean-squared displacement of individual colloids (see Supplementary Note S7 and Supplementary Fig. S14).

The silica colloidal particles, suspended in a $3:7$ (v/v) solution of dimethyl sulfoxide (DMSO, Sigma) and water with the requisite concentration of depletant, $c$, (sodium carboxymethyl cellulose (NaCMC, Sigma)) was introduced into the flow cell via a syringe pump. We performed experiments at six depletion concentrations, $(0.10 \leq c \leq 0.20)$ mg/ml. In all our experiments, the mean separation between nucleating islands, $l$, at maximum island density was observed to be $l \sim (4.8 \pm 1.0) \sigma$ (mean $\pm$ standard deviation) \cite{ganapathy2010direct}. Since $l << L_p$, the possibility of nucleation at one pinned site is independent of others, and therefore, effective interaction between pinned sites can be safely ruled out. Further, to ensure the independence of observed phenomena irrespective of random configurations of pinned particles, two experiments were conducted at each $c$ (except $c= 0.20$ mg/ml), employing distinct random configurations of pinned particles while maintaining similar $L_p$, both under similar experimental conditions.

The video microscopy of the vapor deposition process was observed using an optical microscope (Leica DMI8, Camera: Hamamatsu Orca Flash4) using a 100x oil immersion objective. The typical field-of-view was $\sim 114 \times 114$ $\mu \text{m}^2$. Image acquisition was conducted at varying frame rates: initially at 10 frames per second (fps) for the first 30 minutes, transitioning to 4 fps for the subsequent 90 minutes, and eventually settling at 2 fps for the final 90 minutes. During the advanced stage of crystal growth, typically occurring $\sim 3.5$ hours into the experiments with significant monolayer coverage, an additional 15 minutes of data was captured at 15 fps. The particles were tracked using standard algorithms \cite{crocker1996methods}, clusters were identified using a distance-based cut-off algorithm, and analyses were performed using in-house developed codes in Python. The dynamic spatial resolution in our experiment was $ \sim 25$ nm.

\subsubsection*{Experiments to study normal modes of vibration}
To achieve crystallites without particles in the second layer, necessary for data reported in Fig. \ref{fig:Figure4}, we implemented a modified experimental approach. These experiments were conducted within cylindrical experimental cells (dimension [height $\times$ diameter]: $\sim 9$ mm $\times \sim 7$ mm). The increased height of the cell enabled us to attain a low sedimentation flux, specifically $F = (1.10 \pm 0.50) \times 10^{-5}$ monolayers/s, for these experiments. 

As discussed in the main text, to address the challenge posed by the infrequent occurrence of pinned particles away from the edges of the crystallites, we employed a systematic approach, scanning the entire experimental cell. However, this approach made tracking and identification of the pinned particles challenging. To address this, we designed a method to \textit{in-situ} distinguish between pinned and free particles by utilizing two types of particles with distinct refractive indices, resulting in a noticeable brightness contrast when viewed through an optical microscope. Specifically, the bottom glass coverslip of the experimental cell was spin-coated with fluorescent polystyrene spheres (Sigma, diameter:  1.0 $\mu$m, size polydispersity: 3\%). These spheres served as the pinned sites. Meanwhile, silica colloids (Sigma, diameter: 1.0 $\mu$m, size polydispersity: 3\%) were utilized for the vapor deposition process. The experiments were performed at $c = 0.12$ mg/ml. Once the desired regions-of-interest containing a pinned particle away from the crystallite edges were identified, images were captured at 20 fps for 20 minutes.

\subsection*{Theoretical Model}
In order to capture the essential experimental phenomenology, we develop a minimal theoretical model to describe the system. We model the colloids as hard spheres hopping on a two-dimensional triangular lattice. The choice of the lattice is guided by the observation that its coordination number ($=6$) matches the average coordination number of a Delaunay triangulated random graph. The dynamics of the system are modelled within a Markov matrix-based framework, where the total number of particles is conserved. This assumption aligns well with the conditions of the vapor deposition experiments under consideration, characterized by slow influx rates and sluggish diffusivity of the particles on the substrate. This also allows us to calculate thermodynamic quantities, as discussed in more detail below.

\subsubsection*{Markov Process: Transitions between the configurations}\label{sec_markov}
Let us consider a triangular lattice with $L$ sites, indexed by $i=1,2,...,L$ and each of these sites can either be occupied by a particle or remain unoccupied. Thus, total number of possible configurations is $\Omega=2^{L}$. We consider transitions between these configurations as Markov processes \cite{feller68_prob,livi_Politi_2017}, where probability of having $j^{\text{th}}$ configuration at time $t$ is $q_{j}(t)$ (where $0 \leq q_{j} (t) \leq 1$), such that $\sum_{j=1}^{\Omega} q_{j}(t)=1$. Thus, the rate of change of probabilities $Q(t)=\begin{pmatrix}
q_{1}(t) & q_{2}(t) & .&. & . & q_{\Omega}(t)
\end{pmatrix}^{T}$ is determined by
\begin{equation}
\frac{d Q(t)}{dt}=M Q(t), 
\end{equation}
with $M$ being $\Omega \times \Omega$ Markov matrix. If $w_{jk}$ is transition rate from $k^{\text{th}}$ configuration to $j^{\text{th}}$ configuration, then $M_{jk}=w_{jk}$ for $j\neq k$ and $M_{kk}=-\sum_{j \neq k} w_{jk}$, such that $\sum_{j} M_{jk}= \sum_{j \neq k} M_{jk} + M_{kk}=0$.

Thus, the row vector $Q_{0}^{T}=\begin{pmatrix}
\frac{1}{\Omega} &\frac{1}{\Omega}& . & . &. & \frac{1}{\Omega}
\end{pmatrix}$
having equal probabilities for all configurations is left eigenvector of $M$ with eigenvalue zero, i.e.,
\begin{equation}
Q_{0}^{T} M=0,
\end{equation}
while the column vector $Q_{\rm{steady}}$ composed of probabilities for steady-state configurations satisfies
\begin{equation}
\frac{d Q_{\rm{steady}}}{dt}=M Q_{\rm{steady}} =0,
\end{equation}
implying $Q_{\rm{steady}}$ being right eigenvector of $M$ with zero eigenvalue. In general, $M$ is not a symmetric matrix (i.e. $M \neq M^{T}$) and thus $Q_{\rm{steady}}$ usually differs from $Q_{0}$. However, if $M$ is a symmetric matrix (i.e. $M=M^{T}$), 
$Q_{\rm{steady}}=Q_{0}$.

Thus, for a symmetric Markov matrix, a steady state corresponds to equal probabilities for all configurations. The steady state of the system, with equal probability of all configurations, implies an effective temperature $T \rightarrow \infty$ such that the corresponding partition function is ${\cal Z} = \Omega$ and the thermodynamic entropy is $S = k_B \ln \Omega$. This approach guarantees that only the entropic contribution of the system is included in the steady state.

Now, we assume that the transitions between the configurations occur due to the exchange (hopping) of particles between occupied and unoccupied nearest neighboring sites in the lattice. Using the analogy with spin-$\frac{1}{2}$, each occupied (unoccupied) site can be interpreted as an up (down) spin state. This analogy enables us to express Markov matrix as:
\begin{equation}\label{eq_M_spin}
M= w \sum_{\langle jk \rangle} (\sigma_{j}^{+} \sigma_{k}^{-} + \sigma_{k}^{+} \sigma_{j}^{-})(\sigma_{j}^{z}-\sigma_{k}^{z})^{2} - w \sum_{\langle jk \rangle} (\sigma_{j}^{z}-\sigma_{k}^{z})^{2},
\end{equation}
where $\sigma_{j}^{\pm}=\frac{1}{2}(\sigma_{j}^{x} \pm i \sigma_{j}^{y})$
with $\sigma_{j}^{x},\sigma_{j}^{y}, \sigma_{j}^{z}$ being Pauli matrices at $j^{\text{th}}$ site and $\langle .. \rangle$ denotes nearest neighboring sites. Here, the hopping (exchange) process with strength $w$ between any two nearest neighbouring sites is allowed only if one of these two sites is unoccupied and the other is occupied, which is consistent with the hard-sphere assumption of the particles. In terms of spins, this exchange process can occur between the nearest neighbouring sites with opposite spin-states (up and down), thus resulting in the term $(\sigma_{j}^{z}-\sigma_{k}^{z})^{2}$ for the exchange process (see first term in Eq.~\eqref{eq_M_spin}). The last term on the right hand side of Eq.~\eqref{eq_M_spin} ensures the property $\sum_{j} M_{jk}=0$ of Markov matrix. Using the relations $\sigma_{j}^{+}\sigma_{j}^{z}=-\sigma_{j}^{+}$ and $\sigma_{j}^{-}\sigma_{j}^{z}=\sigma_{j}^{-}$, Eq.~\eqref{eq_M_spin} can be re-written as
\begin{equation}\label{eq_heisenberg}
M= 2 w \sum_{\langle jk \rangle} (\sigma_{j}^{x} \sigma_{k}^{x} + \sigma_{j}^{y} \sigma_{k}^{y}+ \sigma_{j}^{z} \sigma_{k}^{z} ) + {\rm{constant}},
\end{equation}
which resembles the Hamiltonian of spin-$\frac{1}{2}$ Heisenberg model. Here, the ground state of the ferromagnetic Heisenberg Hamiltonian $-M$ (where $w>0$) corresponds to the steady state 
for Markov process discussed above. As $M$ in Eq.~\eqref{eq_heisenberg} is a symmetric matrix, all the configurations are equally probable in the steady state.

Now, let us discuss about Markov matrix in presence of pinning of particles. When a particle is pinned at a fixed site, that particle cannot participate in hopping process. Thus, pinning of a particle corresponds to fixing the spin to an up-spin state and pinning of $L_{\rm{pin}}$ sites leads to reduction of effective number of configurations to $2^{L-L_{\rm{pin}}}$. In this situation, the Markov matrix can still be written in same form as in Eq.~\eqref{eq_M_spin} where the summation (in Eq.~\eqref{eq_M_spin}) now runs over all pairs of nearest neighbouring sites excluding the pairs having a pinned site.

As the number of configurations $\Omega=2^{L}$ increases exponentially with system-size $L$, exploring all steady-state configurations in numerical computation becomes impossible. Therefore, we resort to a different numerical scheme, as discussed in the following section, where configurations are randomly chosen and analysed among all equally probable steady-state configurations (see Supplementary Note S5).


\bibliography{bibliography} 

\begin{thebibliography}{10}
\providecommand{\url}[1]{\texttt{#1}}
\expandafter\ifx\csname urlstyle\endcsname\relax
  \providecommand{\doi}[1]{doi:\discretionary{}{}{}#1}\else
  \providecommand{\doi}{doi:\discretionary{}{}{}\begingroup
  \urlstyle{rm}\Url}\fi

\bibitem{granasy2003growth}
L.~Granasy, \emph{et~al.}, Growth of'dizzy dendrites' in a random field of
  foreign particles. \emph{Nat. Mater.} \textbf{2}~(2), 92 (2003).

\bibitem{musevic2006two}
I.~Musevic, M.~Skarabot, U.~Tkalec, M.~Ravnik, S.~Zumer, Two-dimensional
  nematic colloidal crystals self-assembled by topological defects.
  \emph{Science} \textbf{313}~(5789), 954 (2006).

\bibitem{zhong1993computer}
W.~Zhong, Y.~Cai, D.~Tomanek, Computer simulation of hydrogen embrittlement in
  metals. \emph{Nature} \textbf{362}~(6419), 435 (1993).

\bibitem{kermode2013macroscopic}
J.~R. Kermode, \emph{et~al.}, Macroscopic scattering of cracks initiated at
  single impurity atoms. \emph{Nat. Commun.} \textbf{4}~(1), 2441 (2013).

\bibitem{delattre2009noisy}
T.~Delattre, \emph{et~al.}, Noisy kondo impurities. \emph{Nat. Phys.}
  \textbf{5}~(3), 208 (2009).

\bibitem{mocatta2011heavily}
D.~Mocatta, \emph{et~al.}, Heavily doped semiconductor nanocrystal quantum
  dots. \emph{Science} \textbf{332}~(6025), 77 (2011).

\bibitem{cacciuto2004onset}
A.~Cacciuto, S.~Auer, D.~Frenkel, Onset of heterogeneous crystal nucleation in
  colloidal suspensions. \emph{Nature} \textbf{428}~(6981), 404 (2004).

\bibitem{navrotsky2004energetic}
A.~Navrotsky, Energetic clues to pathways to biomineralization: Precursors,
  clusters, and nanoparticles. \emph{Proc. Natl. Acad. Sci. USA}
  \textbf{101}~(33), 12096 (2004).

\bibitem{elhadj2006role}
S.~Elhadj, J.~De~Yoreo, J.~Hoyer, P.~Dove, Role of molecular charge and
  hydrophilicity in regulating the kinetics of crystal growth. \emph{Proc.
  Natl. Acad. Sci. USA} \textbf{103}~(51), 19237 (2006).

\bibitem{de2005colloidal}
V.~W. De~Villeneuve, \emph{et~al.}, Colloidal hard-sphere crystal growth
  frustrated by large spherical impurities. \emph{Science} \textbf{309}~(5738),
  1231 (2005).

\bibitem{allahyarov2015crystallization}
E.~Allahyarov, K.~Sandomirski, S.~U. Egelhaaf, H.~L{\"o}wen, Crystallization
  seeds favour crystallization only during initial growth. \emph{Nat. Commun.}
  \textbf{6}~(1), 7110 (2015).

\bibitem{ma2021immobilized}
L.~Ma, \emph{et~al.}, Immobilized precursor particle driven growth of
  centimeter-sized MoTe\(_2\) monolayer. \emph{J. Am. Chem. Soc.}
  \textbf{143}~(33), 13314 (2021).

\bibitem{hong2020chemical}
Y.-L. Hong, \emph{et~al.}, Chemical vapor deposition of layered two-dimensional
  MoSi\(_2\)N\(_4\) materials. \emph{Science} \textbf{369}~(6504), 670 (2020).

\bibitem{ma2020antagonistic}
W.~Ma, J.~F. Lutsko, J.~D. Rimer, P.~G. Vekilov, Antagonistic cooperativity
  between crystal growth modifiers. \emph{Nature} \textbf{577}~(7791), 497
  (2020).

\bibitem{katsuno2011effect}
H.~Katsuno, K.~Katsuno, M.~Sato, Effect of immobile impurities on
  two-dimensional nucleation. \emph{Physical Review E} \textbf{84}~(2), 021605
  (2011).

\bibitem{kellogg1994direct}
G.~Kellogg, Direct observation of substitutional-atom trapping on a metal
  surface. \emph{Phys. Rev. Lett.} \textbf{72}~(11), 1662 (1994).

\bibitem{liu1995submonolayer}
S.~Liu, L.~B{\"o}nig, J.~Detch, H.~Metiu, Submonolayer growth with repulsive
  impurities: Island density scaling with anomalous diffusion. \emph{Phys. Rev.
  Lett.} \textbf{74}~(22), 4495 (1995).

\bibitem{li2016assembly}
B.~Li, D.~Zhou, Y.~Han, Assembly and phase transitions of colloidal crystals.
  \emph{Nat. Rev. Mater.} \textbf{1}~(2), 1 (2016).

\bibitem{manoharan2015colloidal}
V.~N. Manoharan, Colloidal matter: Packing, geometry, and entropy.
  \emph{Science} \textbf{349}~(6251), 1253751 (2015).

\bibitem{van2008colloidal}
S.~van Teeffelen, C.~N. Likos, H.~L{\"o}wen, Colloidal crystal growth at
  externally imposed nucleation clusters. \emph{Phys. Rev. Lett.}
  \textbf{100}~(10), 108302 (2008).

\bibitem{ganapathy2010direct}
R.~Ganapathy, M.~R. Buckley, S.~J. Gerbode, I.~Cohen, Direct measurements of
  island growth and step-edge barriers in colloidal epitaxy. \emph{Science}
  \textbf{327}~(5964), 445 (2010).

\bibitem{meng2010free}
G.~Meng, N.~Arkus, M.~P. Brenner, V.~N. Manoharan, The free-energy landscape of
  clusters of attractive hard spheres. \emph{Science} \textbf{327}~(5965), 560
  (2010).

\bibitem{perry2015two}
R.~W. Perry, M.~C. Holmes-Cerfon, M.~P. Brenner, V.~N. Manoharan,
  Two-dimensional clusters of colloidal spheres: Ground states, excited states,
  and structural rearrangements. \emph{Phys. Rev. Lett.} \textbf{114}~(22),
  228301 (2015).

\bibitem{juttner2018vf2++}
A.~J{\"u}ttner, P.~Madarasi, VF2++—An improved subgraph isomorphism
  algorithm. \emph{Discrete Appl. Math.} \textbf{242}, 69 (2018).

\bibitem{cordella2001improved}
L.~P. Cordella, P.~Foggia, C.~Sansone, M.~Vento, An improved algorithm for
  matching large graphs, in \emph{3rd IAPR-TC15 workshop on graph-based
  representations in pattern recognition} (Citeseer) (2001), p. 149.

\bibitem{hagberg2008exploring}
A.~A. Hagberg, D.~A. Schult, P.~J. Swart, Exploring network structure,
  dynamics, and function using NetworkX, in \emph{Proceedings of the 7th Python
  in Science Conference (SciPy2008)}, G.~Varoquaux, T.~Vaught, J.~Millman, Eds.
  (Pasadena, CA USA) (2008), p.~11.

\bibitem{asakura1958interaction}
S.~Asakura, F.~Oosawa, Interaction between particles suspended in solutions of
  macromolecules. \emph{J. Polym. Sci.} \textbf{33}~(126), 183 (1958).

\bibitem{LAMMPS}
A.~P. Thompson, \emph{et~al.}, {LAMMPS} - a flexible simulation tool for
  particle-based materials modeling at the atomic, meso, and continuum scales.
  \emph{Comp. Phys. Comm.} \textbf{271}, 108171 (2022).

\bibitem{feller68_prob}
W.~Feller, \emph{An introduction to probability theory and its applications,
  Volume 1} (John Wiley \& Sons) (1968).

\bibitem{livi_Politi_2017}
R.~Livi, P.~Politi, \emph{Nonequilibrium Statistical Physics: A Modern
  Perspective} (Cambridge University Press) (2017).

\bibitem{essam_1980}
J.~W. Essam, Percolation theory. \emph{Reports on Progress in Physics}
  \textbf{43}~(7), 833 (1980).

\bibitem{stauffer2018introduction}
D.~Stauffer, A.~Aharony, \emph{Introduction to percolation theory} (Taylor \&
  Francis) (2018).

\bibitem{sykesI_1976}
M.~F. Sykes, M.~Glen, Percolation processes in two dimensions. I. Low-density
  series expansions. \emph{Journal of Physics A: Mathematical and General}
  \textbf{9}~(1), 87 (1976).

\bibitem{doi:10.1126/science.1128649}
J.~R. Savage, D.~W. Blair, A.~J. Levine, R.~A. Guyer, A.~D. Dinsmore, Imaging
  the Sublimation Dynamics of Colloidal Crystallites. \emph{Science}
  \textbf{314}~(5800), 795 (2006).

\bibitem{swallen2007organic}
S.~F. Swallen, \emph{et~al.}, Organic glasses with exceptional thermodynamic
  and kinetic stability. \emph{Science} \textbf{315}~(5810), 353 (2007).

\bibitem{zhu2011surface}
L.~Zhu, \emph{et~al.}, Surface self-diffusion of an organic glass. \emph{Phys.
  Rev. Lett.} \textbf{106}~(25), 256103 (2011).

\bibitem{chen2010low}
K.~Chen, \emph{et~al.}, Low-frequency vibrations of soft colloidal glasses.
  \emph{Phys. Rev. Lett.} \textbf{105}~(2), 025501 (2010).

\bibitem{melio2024soft}
J.~Melio, S.~E. Henkes, D.~J. Kraft, Soft and stiff normal modes in floppy
  colloidal square lattices. \emph{Phys. Rev. Lett.} \textbf{132}~(7), 078202
  (2024).

\bibitem{arsenault2007photonic}
A.~C. Arsenault, D.~P. Puzzo, I.~Manners, G.~A. Ozin, Photonic-crystal
  full-colour displays. \emph{Nat. Photonics} \textbf{1}~(8), 468 (2007).

\bibitem{courty2005vibrational}
A.~Courty, A.~Mermet, P.~Albouy, E.~Duval, M.~Pileni, Vibrational coherence of
  self-organized silver nanocrystals in fcc supra-crystals. \emph{Nat. Mater.}
  \textbf{4}~(5), 395 (2005).

\bibitem{stober1968controlled}
W.~St{\"o}ber, A.~Fink, E.~Bohn, Controlled growth of monodisperse silica
  spheres in the micron size range. \emph{J. Colloid Interface Sci.}
  \textbf{26}~(1), 62 (1968).

\bibitem{zhang2009hollow}
L.~Zhang, \emph{et~al.}, Hollow silica spheres: synthesis and mechanical
  properties. \emph{Langmuir} \textbf{25}~(5), 2711 (2009).

\bibitem{crocker1996methods}
J.~C. Crocker, D.~G. Grier, Methods of digital video microscopy for colloidal
  studies. \emph{J. Colloid Interface Sci.} \textbf{179}~(1), 298 (1996).

\end{thebibliography}
\bibliographystyle{sciencemag}


\section*{Acknowledgments}
Authors thank Mahesh M Bandi, Sriram Ramaswamy, Chandan Dasgupta, John C. Crocker, Samudrajit Thapa, Prasanna Venkatesh B., K. Hima Nagamanasa, Anirban Sain, Pinaki Majumdar, Sankalp Nambiar, Sivasurender Chandran and Vinay Vaibhav for useful comments and discussions. The idea for this project originated from preliminary experimental observations made by C.K.M. during his PhD studies in the Ganapathy Group at JNCASR. We acknowledge Rajesh Ganapathy for his comments on this problem. 

\section*{Funding}
We gratefully acknowledge financial support from the Department of Science and Technology (Government of India), INSPIRE fellowship IF200274 (N.H.B.), Prime Minister Research Fellowship (S.M.),  IIT Kanpur Initiation Grant through IITK/PHY/2022010 (A.A.),  IIT Gandhinagar through IP/IITGN/PHY/\\CM/2021/11 (C.K.M.) and the Start-up Research Grant of Science and Engineering Research Board of Government of India through SRG/2021/001077 (C.K.M.).

\section*{Competing interests}
The authors declare that they have no competing interests.

\end{document}